\documentclass[9pt,conference]{IEEEtran}
\usepackage{graphicx,float}
\usepackage{caption}
\usepackage[square, numbers]{natbib}
\usepackage{hyperref}
\usepackage{listings}
\hypersetup{colorlinks,linkcolor=blue,filecolor=blue,urlcolor=blue,citecolor=blue}
\def\docroot{.}
\graphicspath{{\docroot/Figures/}}

\title{ESP32: QEMU Emulation within a Docker Container}

\author{
    \IEEEauthorblockN{Michael Howard P.Eng}
    \IEEEauthorblockA{\textit{Department of Computer Science} \\
        \textit{Maseeh College of Engineering and Computer Science }\\
        Portland State University \\
        email: mihoward@pdx.edu
    }
\and
    \IEEEauthorblockN{Dr. R. Bruce Irvin}
    \IEEEauthorblockA{\textit{Department of Computer Science} \\
        \textit{Maseeh College of Engineering and Computer Science }\\
        Portland State University \\
        email: rbi@pdx.edu
    }
}

\begin{document}
\definecolor{codegreen}{rgb}{0,0.6,0}
\definecolor{codegray}{rgb}{0.5,0.5,0.5}
\definecolor{codepurple}{rgb}{0.58,0,0.82}
\definecolor{backcolour}{rgb}{0.95,0.95,0.92}
\lstdefinestyle{mystyle}{
    backgroundcolor=\color{backcolour},   
    commentstyle=\color{codegreen},
    keywordstyle=\color{magenta},
    stringstyle=\color{codepurple},
    basicstyle=\ttfamily\footnotesize,
    breakatwhitespace=false,         
    breaklines=true,                 
    captionpos=b,                    
    keepspaces=true,                 
    numbers=none,                    
    numbersep=5pt,                  
    showspaces=false,                
    showstringspaces=false,
    showtabs=false,                  
    tabsize=2
}
\lstset{style=mystyle, inputpath=\docroot/Listings}

\maketitle
\begin{abstract}
    The ESP32 is a popular microcontroller from Espressif that can be used in many embedded applications.  Robotic joints, smart car chargers, beer vat agitators and automated bread mixers are a few examples where this system-on-a-chip excels.  It is cheap to buy and has a number of vendors providing low-cost development board kits that come with the microcontroller and many external connection points with peripherals.

    There is a large software ecosystem for the ESP32.  Espressif maintains an SDK containing many C-language sample projects providing a starting point for a huge variety of software services and I/O needs.  Third party projects provide additional sample code as well as support for other programming languages.  For example, MicroPython is a mature project with sample code and officially supported by Espressif.  The SDK provides tools to not just build an application but also merge a flash image, flash to the microcontroller and monitor the output.
    
    Is it possible to build the ESP32 load and emulate on another host OS?  This paper explores the QEMU emulator and its ability to emulate the ethernet interface for the guest OS.  Additionally, we look into the concept of containerizing the entire emulator and ESP32 load package such that a microcontroller flash image can successfully run with a one-step deployment of a Docker container.
\end{abstract}\label{abstract}

\section{Introduction}\label{sec:intro}
The ESP32 is a microcontroller developed by Espressif Systems Co. \cite{esp2023}, a semiconductor company headquartered in Shanghai, China.  The ESP32 provides a low-cost, low-power and reasonably performant all-in-one hardware package that is ideally suited to Internet of Things (IoT) applications.  The RISC processor is a 32-bit Xtensa core developed by Cadence Design Systems \cite{cadence2023}.  It is packaged as a system-on-a-chip (SoC) with Bluetooth, Wi-Fi and general purpose input output (GPIO) capabilities built-in.

In addition to the suite of microcontrollers, Espressif also supports a rich software ecosystem.  Developers around the world contribute to an open source project that provide a software development kit (SDK) for each microcontroller flavor.  The SDK provides a complete build system that utilize Python scripting on top of cmake C-language build projects.  Device drivers, a real-time operating system (OS) called FreeRTOS and a suite of software component projects are all included.  The component projects provide an excellent starting point for developers to expand their own software requirements.

QEMU \cite{qemu2023} is an emulator able to run guest OS and application binaries on a host operating system.  For this project, the overall goal is to see how QEMU can be used to emulate an ESP32 application image on a macOS host system.  

\subsection{Goals}\label{sub:goals}
At the completion of the project, the following target goals will be accomplished:
\begin{enumerate}
    \item Build an ESP32 target load containing the OS, device drivers and a simple HTTP application such as a web server that uses the TCP/IP stack.\label{goal:build}
    \item Execute this load on native hardware to first ensure it is functional.\label{goal:exec}
    \item Custom build QEMU for macOS (with ESP32 support) and modify as needed to run the target load.\label{goal:qemu}
    \item  Develop a Docker container around a QEMU tool chain.  A container will provide an isolated environment for all dependencies and is preferable to running natively.\label{goal:container}
    \item Generate some minimal HTTP traffic between the application and the external world.\label{goal:traffic}
    \item If working, trace through the call stack(s) as much as possible and determine how the emulation is being performed.\label{goal:trace}
    \item Detail and report on the experiments and system architecture.\label{goal:report}\\    
\end{enumerate}

The overall goal as detailed above is motivated by a desire to easily emulate the ESP32.  This microcontroller is being used as an edge device in the author's energy auction research.  As such, the need to easily deploy and destroy multiple containers containing ESP32 loads (with emulators) will facilitate load testing and easy automation for functional tests.  Both the Capstone project and follow-on research will benefit from being able to build the ESP32 load directly into a container and then immediately deploy multiple instances, each with a unique identifier, on either a local workstation or Cloud container service.  Additionally, the knowledge and understanding gained through digging into the QEMU architecture and APIs will  assist with the development and debugging of the edge device application. 

The system architecture is outlined in Section \ref{sec:system}.  The experiments performed and corresponding analysis are detailed in Sections \ref{sec:experiments} and \ref{sec:analysis} respectively.  The final discussion and conclusions are captured in Section \ref{sec:conclusion}.  
\section{System Architecture}\label{sec:system}

The relevant system architecture for this paper primarily consists of the ESP32 microcontroller, its SDK, the QEMU emulator and Docker containers.  Visual Studio Code provides extensions that interface both to the ESP32 SDK and the Docker container management.  While the VSCode integrated development environment (IDE) does provide convenience features used during the experiments, they will only be mentioned in passing without going in depth in their implementation.

\subsection{ESP32 microcontroller}\label{sub:esp32}
\begin{figure}[H]
    \centerline{\includegraphics[width=1.0\linewidth, keepaspectratio]{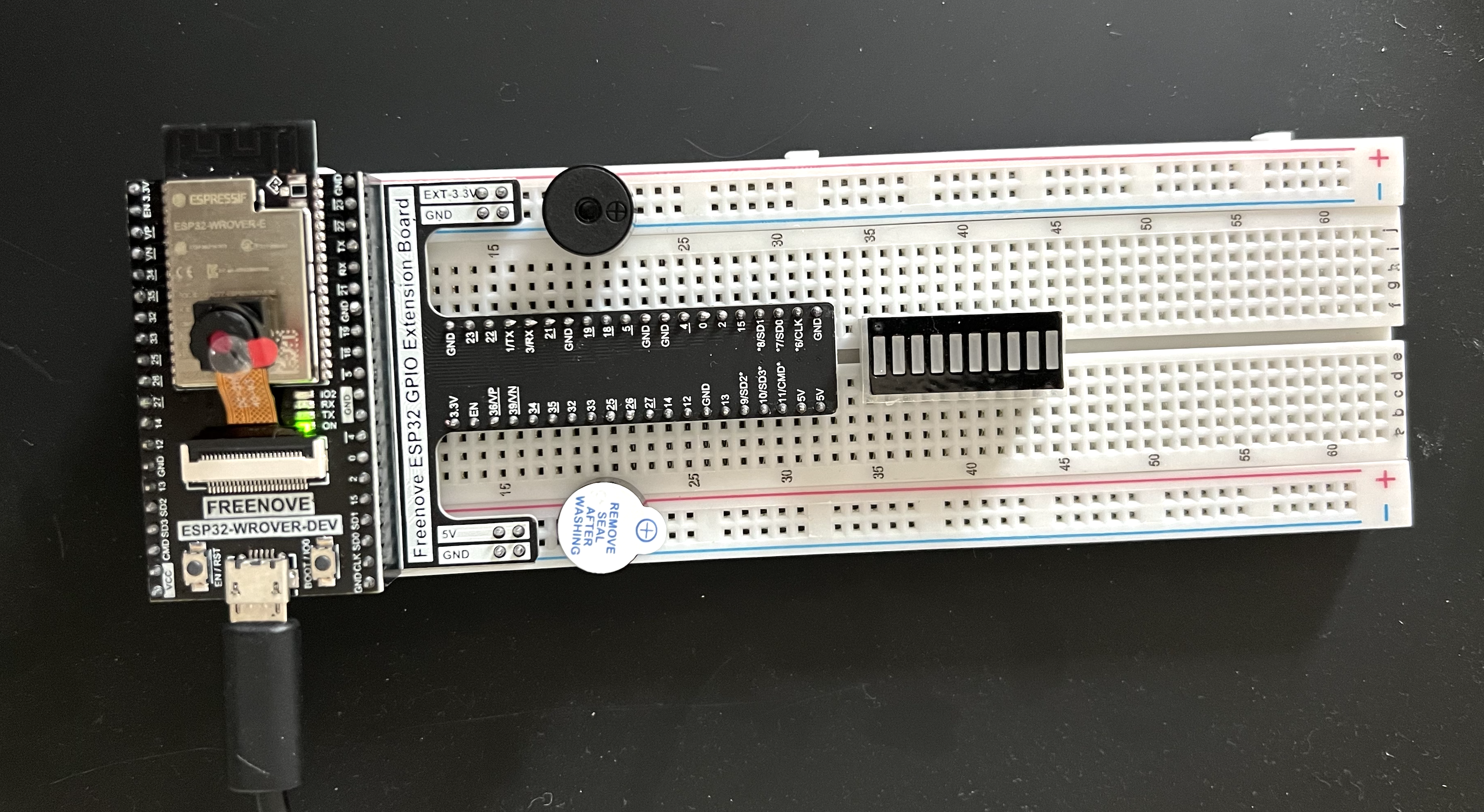}}
    \captionsetup{width=.8\linewidth}
    \caption{FreeNove ESP32 development board containing the core SoC with some sample I/O devices such as LEDs and speaker.}\label{fig:esp_hw}
\end{figure}

The core component of this paper is the ESP32 microcontroller as this is the desired platform to both emulate and virtualize. Figure \ref{fig:esp_hw} shows the hardware development kit purchased for this experiment.  A micro USB connector provides both power and data.  The SoC chip itself can be purchased very cheaply on its own.  The dev kit as shown adds a few extra (unnecessary) peripherals such as a breadboard, GPIO breakout pins, LEDs, speakers and a camera.  The entire package was purchased for \$26.00 on Amazon which also shipped with a set of wires and resistors.  

\begin{figure}[h]
    \centerline{\includegraphics[width=0.9\linewidth, keepaspectratio]{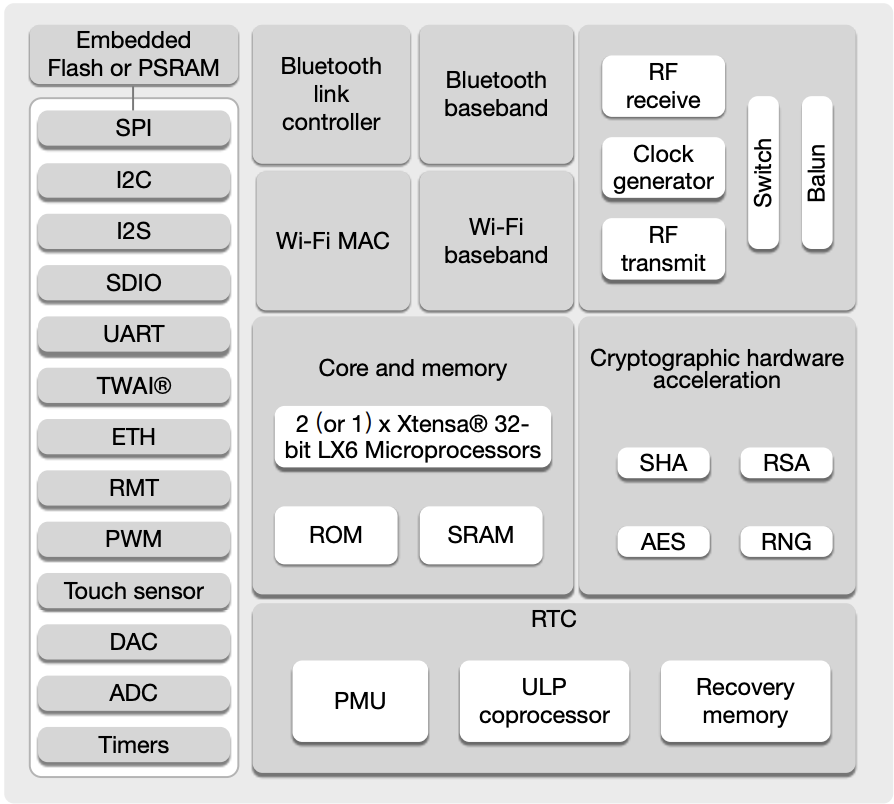}}
    \captionsetup{width=.8\linewidth}
    \caption{The ESP32 block diagram showing the dual Xtensa cores with a wide variety of I/O controllers built in to the chip.}\label{fig:esp_block}   
\end{figure}

The ESP32 chip is the silver component in the upper-left corner of Figure \ref{fig:esp_hw}. Its functional block diagram is shown in Figure \ref{fig:esp_block}. This diagram shows the richness of the SoC with dual Xtensa CPU cores, Wi-Fi, Bluetooth, RAM, flash memory and a large variety of I/O controllers. 

\subsection{Espressif ESP-IDF}\label{sub:esp_idf}
In addition to providing the ESP32 hardware itself, Espressif also maintains an open-source project for a complete SDK called the ESP integrated development framework (ESP-IDF).  Version 5.0 is the current release and publicly available on GitHub \cite{esp_git}.  ESP-IDF provides a suite of Python scripts that can build the application, flash to the remote ESP32 chip and monitor its stdout via the USB port powering the device.

ESP-IDF also consists of large set of cmake \cite{cmake} build projects.  Source code is C-language only and includes the FreeRTOS \cite{freertos} kernel, drivers, support libraries and many sample application projects.  The application developed for this paper is one of the sample projects that provides an HTTP server listening on port 80.  It provides a simple response message to a request message with the \emph{/hello} context.  

\begin{figure}[h]
    \centerline{\includegraphics[width=0.7\linewidth, keepaspectratio]{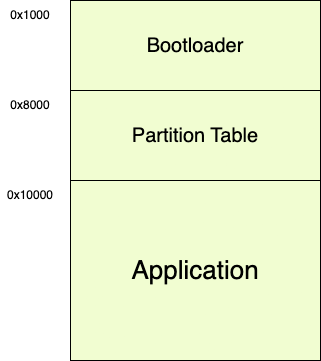}}
    \captionsetup{width=.8\linewidth}
    \caption{The ESP32 flash image layout including bootloader, partition table and application.}\label{fig:esp_flash}   
\end{figure}

Building the application provides a linked binary containing the application, kernel and libraries all compiled for the Xtensa instruction set.  However, this monolith cannot run on the ESP32 target without providing some additional components for the flash device.  Both a bootloader and a partition table must be present in order for the microcontroller to bootstrap the application load.  ESP-IDF provides tools to assemble (merge) a final binary flash image.  The layout is shown in Figure \ref{fig:esp_flash}.  This merged binary is now able to be written to the target flash and booted.  It is ready for emulation as well. 

\subsection{QEMU}\label{sub:qemu}
QEMU is a system emulator that provides a virtual model of a machine to run a guest OS.  CPU, memory and devices are all part of the emulation.  While the vanilla code does support the Xtensa processor, the ESP32 microcontroller currently requires a fork that is maintained by Espressif.  The source code project is available on GitHub \cite{esp_qemu}.  This source project has to be built in order to run QEMU and ESP32 on macOS.  There are no pre-built versions hosted.

QEMU is a user space application that requires an accelerator (hypervisor) in the host kernel.  However, this custom build utilizes Tiny Code Generator (TCG) which is pure emulation.  Theoretically, this trades performance for ease of implementation.  Features such as a block and character device are built-in which allow emulation of stdio, files, sockets and TCP networking.

\subsection{Containers}\label{sub:containers}
\begin{figure}[H]
    \centerline{\includegraphics[width=0.7\linewidth, keepaspectratio]{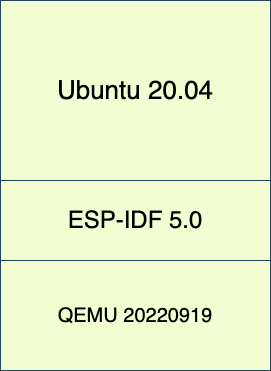}}
    \captionsetup{width=.8\linewidth}
    \caption{The Docker container image containing the Ubuntu base, ESP-IDF SDK toolchain and QEMU emulator.}\label{fig:container_image}   
\end{figure}

The container engine is provided by Docker \cite{docker}.  The purpose of this engine is to provide a virtualization of the file system and configuration while not incurring the overhead of booting a completely separate OS kernel.  The Docker image used starts with an existing Ubuntu 20.04 file system.  The ESP-IDF is cloned into the image as well as a pre-built QEMU from 2022-09-19.  Note that in the case of the container, we can use a pre-built QEMU (Linux binaries) that supports the ESP32.  Espressif has this tarball available in their GitHub repo.  Figure \ref{fig:container_image} shows a coarse outline of the layers in the image.  Each Docker image is composed of layers that can be reused among different images.  Each layer typically maps to a Docker build instruction. 

\section{Experiments}\label{sec:experiments}

The objective of the experiments is to build the ESP32 application, build QEMU, emulate the application via QEMU and containerize both application and QEMU.  Once the setup steps are complete, an HTTP interaction with the application is performed to test and analyze the emulation.

\subsection{Build the ESP32 application}\label{sub:build_app}
The first step is to download and configure the SDK to build the target application.  ESP-IDF is cloned from GitHub \cite{esp_git} and an installation script ensures the necessary binaries and scripts are in the host execution path.  Multiple passes were performed.  However, the final pass took advantage of the VSCode extension which wrapped the underlying clone and install operations.  All build, merge, flash and monitor commands were subsequently run via the \emph{Espressif IDF} extension in VSCode.

After cloning and installing ESP-IDF, the template project selector was used to create a new project based on the simple HTTP server template.  The first pass enabled Wi-Fi through the \begin{quote}CONFIG\_EXAMPLE\_CONNECT\_WIFI=y\end{quote} project configuration.  This repo may be publicly viewed at \url{https://github.com/zemar/esp\_http\_server}.  The application was built, merged into a flash image and flashed to the USB-connected ESP32 target board.  The monitor command then prints all stdout messages from the USB port.

\subsection{Natively run the ESP32 application}\label{sub:native_run}
\begin{figure}[H]
    \centerline{\includegraphics[width=1.0\linewidth, keepaspectratio]{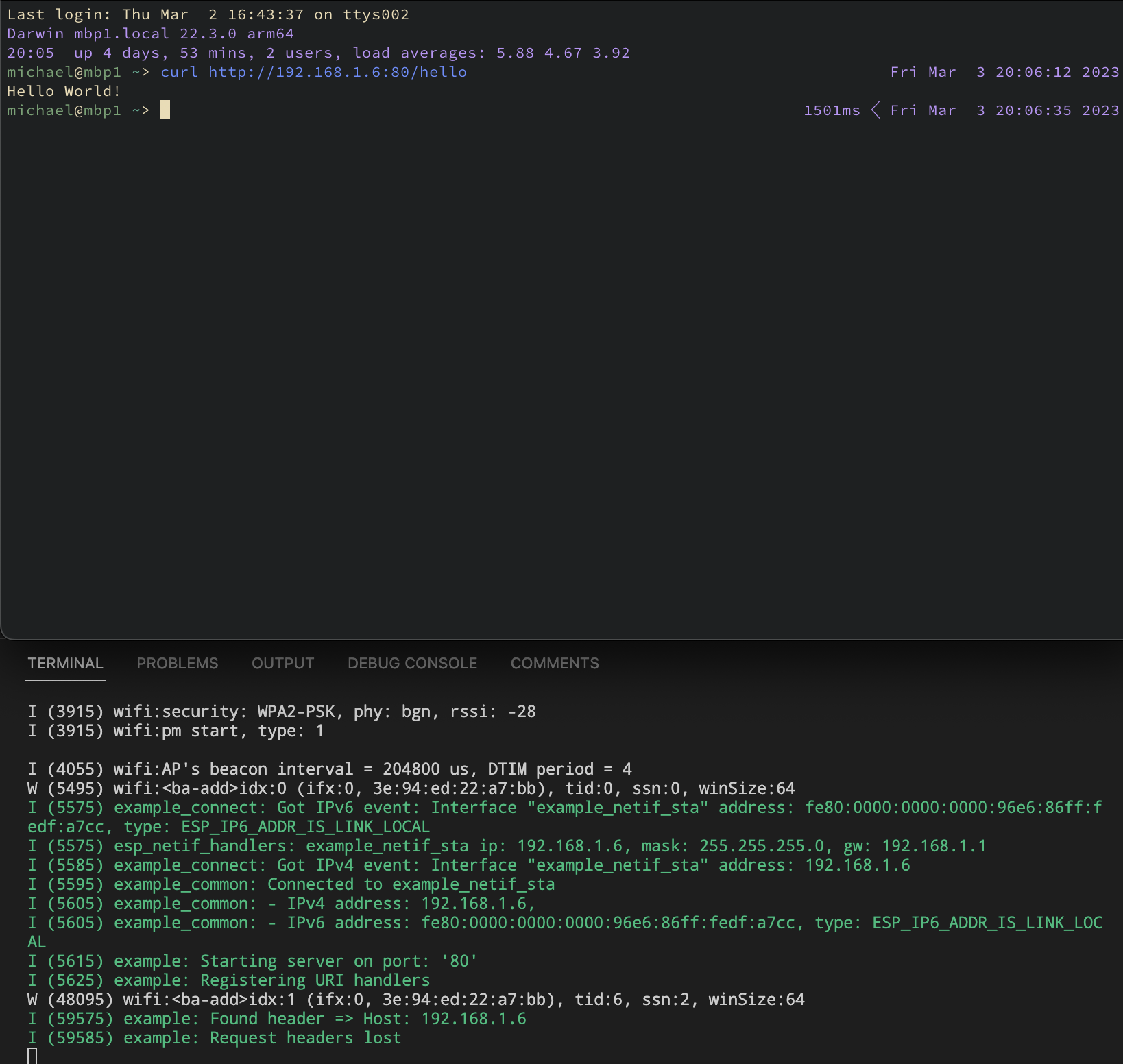}}
    \captionsetup{width=.8\linewidth}
    \caption{Screenshot of host sending a ``/hello'' request with the corresponding ``Hello World!'' response.  Stdout of the ESP32 USB port is also displayed showing the internal handling of the message.}\label{fig:target_response}   
\end{figure}

After loading and monitoring the target board, an HTTP message request was sent from the host.  The correct response message contains ``Hello World!''.  The stdout of the target board is also displayed.  See Figure \ref{fig:target_response} for the actual results.

\subsection{Emulating the ESP32 application}\label{sub:emulate_app}
Since the QEMU fork for ESP32 does not maintain pre-built binaries for macOS, the first step is to clone and build the emulator.  The following steps are used to accomplish this:  

\begin{lstlisting}
    git clone https://github.com/espressif/qemu
    brew install libgcrypt
    ./configure --target-list=xtensa-softmmu \
        --enable-gcrypt --enable-debug \ 
        --enable-sanitizers \
        --disable-strip --disable-user \ 
        --disable-capstone --disable-vnc \
        --disable-sdl --disable-gtk
    ninja -C build
\end{lstlisting}

Note that the target is specified as ``xtensa-softmmu'', thus forcing pure software emulation without requiring Apple's hypervisor library.  
The resulting binary is \textbf{qemu-system-xtensa}.  
However, the ESP32 application is configured for Wi-Fi and there is no pass-through to connect the application network stack to the host Wi-Fi device.  
The solution is to build the ESP32 with an experimental OpenCores Ethernet MAC driver \cite{eth_mac} which provides the \emph{open\_eth} device for configuring the network interface.  This driver is able to pass networking transactions through to the host ethernet interface.  It is configured by setting project options: \begin{quote}CONFIG\_EXAMPLE\_CONNECT\_ETHERNET=y\end{quote} and \begin{quote}CONFIG\_EXAMPLE\_USE\_OPENETH=y\end{quote} and rebuilding.

\begin{figure}[H]
    \centerline{\includegraphics[width=1.0\linewidth, keepaspectratio]{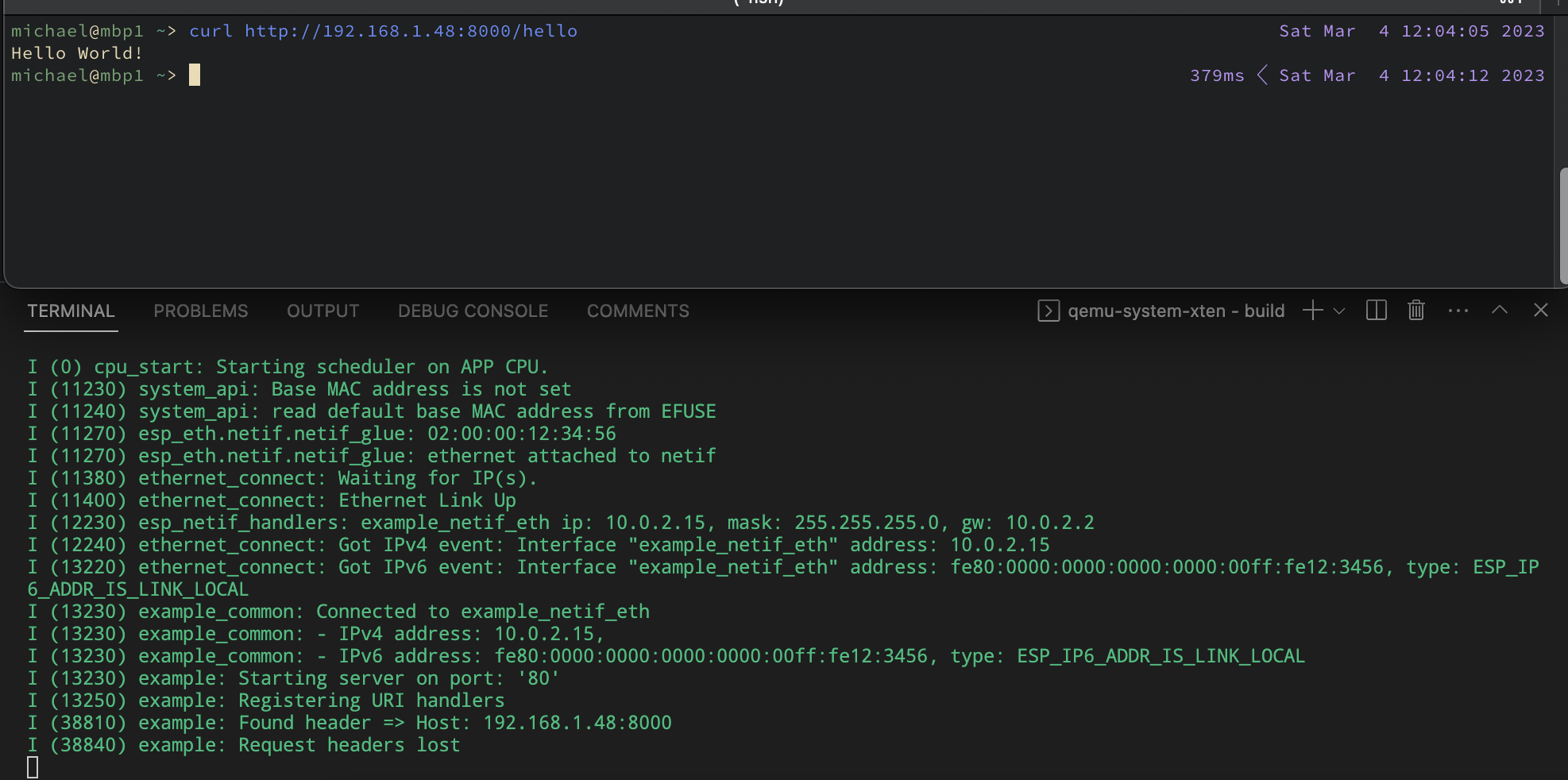}}
    \captionsetup{width=.8\linewidth}
    \caption{Screenshot of host sending a ``/hello'' request with the corresponding ``Hello World!'' response.  Stdout of the QEMU process is also displayed showing the internal handling of the message.}\label{fig:qemu_response}   
\end{figure}

Our custom QEMU now runs (with emulation) the ESP32 flash image using the command:
\begin{lstlisting}
    qemu-system-xtensa -nographic -machine esp32 \ 
        -nic user,model=open_eth,
            id=lo0,hostfwd=tcp::8000-:80 \
        -drive file=merged_qemu.bin,
            if=mtd,format=raw
\end{lstlisting}

Running the above command on our host results in a successful emulation run as shown in Figure \ref{fig:qemu_response}.
The HTTP server is able to bind to 10.0.2.15 on the ``example\_netif\_eth'' interface and the QEMU TCP forwarding allows port 80 in the guest OS to be forwarded to 8000 on the host.  Our HTTP request to port 8000 is then successfully processed with the correct returned response.

\subsection{Containerizing the emulated application}\label{sub:containerize_app}
Building the container image described in Figure \ref{fig:container_image} is straight forward using the Docker engine client and the following command:
\begin{lstlisting}
    docker build -t esp\_qemu .
\end{lstlisting}

With the image built and locally stored, it is run taking the QEMU application command as an argument.  This will run immediately upon launching the container.  Note that the ESP32 flash image, \textbf{merged\_qemu.bin} needs to be present in the container at run time.  A volume mount of the build directory accomplishes this.

The command to run for the experiment:
\begin{lstlisting}
    docker run -it --name esp --rm -p 8000:8000 \ 
        -v $(pwd)/build:/app \
        esp-qemu:latest qemu-system-xtensa \
        -nographic -machine esp32 \ 
        -nic user,model=open\_eth,
            id=lo0,hostfwd=tcp::8000-:80 \
        -drive file=merged\_qemu.bin,
            if=mtd,format=raw
\end{lstlisting}

\begin{figure}[H]
    \centerline{\includegraphics[width=1.0\linewidth, keepaspectratio]{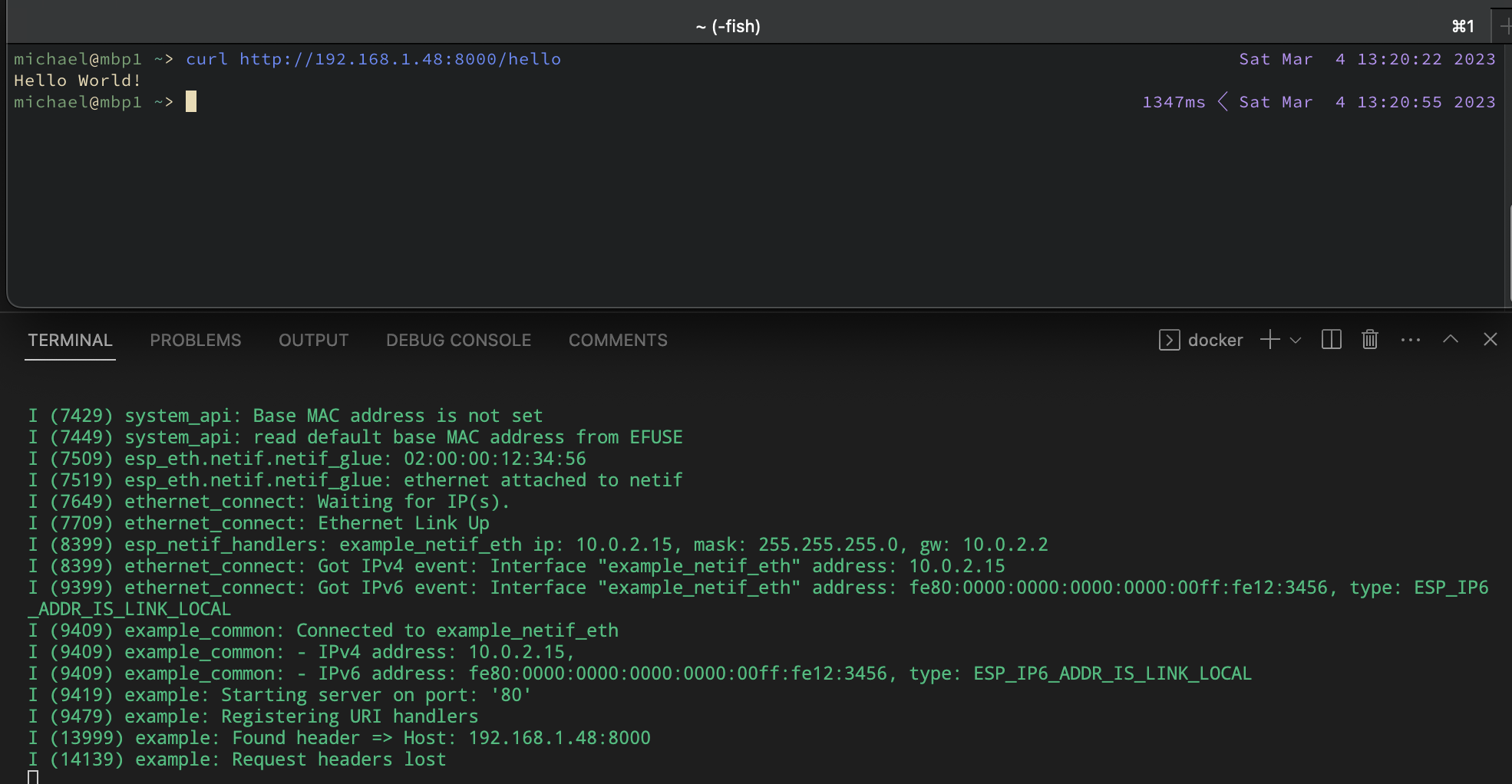}}
    \captionsetup{width=.8\linewidth}
    \caption{Screenshot of host sending a ``/hello'' request with the corresponding ``Hello World!'' response.  Stdout of the Docker container is also displayed showing the QEMU internal handling of the message.}\label{fig:container_response}   
\end{figure}

The above runs the container locally, maps (mounts) the local folder to the container and publishes port 8000 inside the container to the host.  The internal version (inside the container) of the QEMU binary is used to run the mapped \textbf{merged\_qemu.bin}.  As shown in Figure \ref{fig:container_response}, an identical response and set of QEMU messages is generated when virtualized in a container compared to running the emulator natively in macOS.

\section{Analysis}\label{sec:analysis}

Since execution and emulation of the Xtensa instruction set is demonstrated through successfully running the flash binary via QEMU, the primary focus of the analysis is networking.  Emulating the Xtensa is a core feature of the Tiny Code Generator within QEMU.  However, networking does have different possible code paths and configurations as well as a requirement to interface with the host network stack.  Network emulation in QEMU \cite{qemu_net} can take 2 forms: TAP and user mode network stack.  The former adds a virtual network device on the host (called tapN) and can then be configured as a real ethernet card.  The latter was used in this project and will now be described.

\begin{figure}[H]
    \centerline{\includegraphics[width=1.0\linewidth, keepaspectratio]{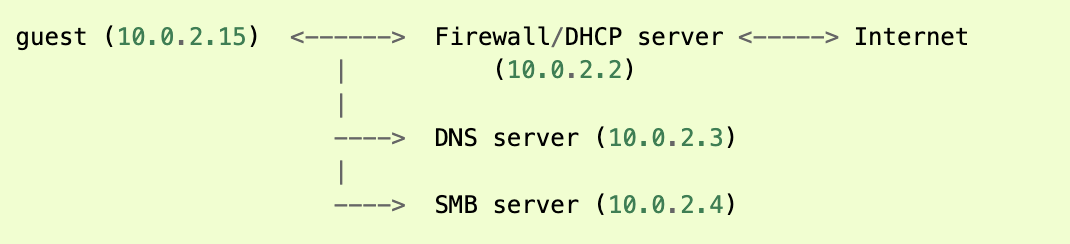}}
    \captionsetup{width=.8\linewidth}
    \caption{The virtual network configuration created by the user mode network stack options.}\label{fig:user_mode_stack}   
\end{figure}

As shown in Section \ref{sub:emulate_app}, the networking interface was configured via launching QEMU with the 
\begin{quote}\textbf{-nic user,model=open\_eth,id=lo0,hostfwd=tcp::8000-:80}\end{quote} options.  
These options configure the user mode network stack without root privileges.  Figure \ref{fig:user_mode_stack} shows the resulting virtual network configuration. The QEMU Virtual Machine (VM) behaves as if it was behind a firewall which blocks all incoming connections.  The DHCP server assigns addresses to the guests starting from 10.0.2.15.

In order for the host to access the IP ports listening on the guest OS, port forwarding must be configured.  The \textbf{hostfwd=tcp::8000-:80} takes care of this by forwarding the guest OS port 80 (our HTTP server) to the host port 8000.  The forwarding is not specific to any one of the host's network interfaces. For instance, both \textbf{lo0} and \textbf{en0} on the host are listening on port 8000.

From Section \ref{sub:emulate_app}, we stated that the ESP32 load was built with support for the OpenCores Ethernet MAC driver \cite{eth_mac}.  This driver, implemented in QEMU, provides the Media Access Control (MAC) layer in the emulator allowing the guest OS to transmit and receive ethernet frames which are subsequently forwarded to the host MAC.  In the QEMU project, \textbf{opencores\_eth.c} implements a set of functions providing the emulated MAC layer interface.  The relevant functions used in transmitting and receiving ethernet frames to the guest OS are:

\begin{itemize}
    \item open\_eth\_desc\_write()
    \item open\_eth\_desc\_read()
    \item open\_eth\_reg\_read()
    \item open\_eth\_reg\_write()
    \item open\_eth\_mii\_read()
    \item open\_eth\_mii\_write()
    \item open\_eth\_start\_xmit()
    \item open\_eth\_receive()
    \item open\_eth\_receive\_desc()
    \item open\_eth\_receive\_mcast()
    \item open\_eth\_update\_irq()
\end{itemize}

\subsection{GDB debugging}\label{sub:gdb}
It is possible to start the QEMU session listening on a debug port.  The ``-s'' option allows this.  Thus, the new launch command on native macOS:

\begin{lstlisting}
    qemu-system-xtensa -nographic -machine esp32 \ 
        -nic user,model=open_eth,
            id=lo0,hostfwd=tcp::8000-:80 \
        -drive file=merged_qemu.bin,
            if=mtd,format=raw \ 
        -s
\end{lstlisting}

Now, we can connect and debug the guest OS and application with: 
\begin{lstlisting}
    xtensa-esp32-elf-gdb esp_http_server.elf \
        -ex "target remote :1234" \
        -ex "monitor system_reset" \
        -ex "tb app_main" -ex "c"
\end{lstlisting}
This GDB session will provide an interactive prompt and will temporarily break at the application entry point.  The FreeRTOS kernel, drivers and HTTP server application is in scope with this debug session.  The \textbf{xtensa-esp32-elf-gdb} is a ESP32-specific build of GDB that is provided through the ESP-IDF SDK.

In theory, GDB stepping through the ESP32 binary will not show anything unique in the emulated session since the guest OS implements a network stack and simply binds to the MAC address at the lower layer.  A more relevant option is to follow the networking code path in the QEMU emulator, specifically the OpenCores Ethernet MAC driver.  Unfortunately, a binary with symbols file (i.e. .elf) is not generated during a custom build which prevents productive GDB stepping through the \textbf{opencores\_eth.c} driver.  This is the entity that provides networking emulation in our experiments.  Also, due to the high transaction rate and asynchronous nature of the network stack operation, GDB stepping will not provide the best analysis of the emulated code path.  A better option is to explore function tracing.

\subsection{QEMU function tracing}\label{sub:tracing}
QEMU provides a function tracing framework which can be enabled at runtime \cite{tracing}. The \begin{quote}\textbf{--trace "open\_eth*"}\end{quote} option will enable tracing on all \textbf{opencores\_eth.c} driver functions.  Thus, we launch QEMU with the following options:

\begin{lstlisting}
    qemu-system-xtensa -nographic -machine esp32 \ 
        -nic user,model=open_eth,
            id=lo0,hostfwd=tcp::8000-:80 \
        -drive file=merged_qemu.bin,
            if=mtd,format=raw \ 
        --trace "open_eth*"
\end{lstlisting}

The new option results in tracing messages for all functions prefixed with ``open\_eth''.  These messages print to stderr.  Log options are also available.  The trace framework is realized by a ``trace()'' function call in each of the subsystem API functions.  The trace function calls outside open\_eth generate a nop such that there is no performance penalty or extraneous messages.

The full stderr trace is too much data to display here.  Listing \ref{lst:trace} is an abbreviation with many in-between open\_eth messages removed.

\begin{lstlisting}[basicstyle=\tiny, numbers=left, label=lst:trace,
  caption={Standard console messages along with trace of open\_eth calls.}]
system_api: Base MAC address is not set
system_api: read default base MAC address from EFUSE
    open_eth_reg_read MAC[0x00] -> 0x0000a000
        .
        .
        .
    open_eth_reg_write MAC[0x44] <- 0x00000200
esp_eth.netif.netif_glue: 02:00:00:12:34:56
esp_eth.netif.netif_glue: ethernet attached to netif
    open_eth_reg_read MAC[0x30] -> 0x00000001
    open_eth_reg_write MAC[0x30] <- 0x00000001
        .
        .
        .
    open_eth_reg_read MAC[0x08] -> 0x00000000
    open_eth_reg_write MAC[0x08] <- 0x00000004
ethernet_connect: Waiting for IP(s).
    open_eth_desc_read DESC[0x0000] -> 0x00002000
    open_eth_desc_write DESC[0x0000] <- 0x015ea000
    open_eth_start_xmit TX: 0x3ffbb964, len: 350, tx_len: 350
    open_eth_receive RX: len: 590
    open_eth_receive_desc RX: 0x3ffba054, len_flags: 0x02524000
    open_eth_update_irq IRQ <- 0x4
        .
        .
        .
    open_eth_desc_write DESC[0x000c] <- 0x3ffba054
ethernet_connect: Ethernet Link Up
    open_eth_desc_read DESC[0x0010] -> 0x0000c000
        .
        .
        .
    open_eth_start_xmit TX: 0x3ffbb964, len: 42, tx_len: 42
    open_eth_desc_write DESC[0x0004] <- 0x3ffbb964

esp_netif_handlers: example_netif_eth ip: 10.0.2.15, mask: 255.255.255.0, gw: 10.0.2.2
ethernet_connect: Got IPv4 event: Interface "example_netif_eth" address: 10.0.2.15
    open_eth_reg_read MAC[0x30] -> 0x00001001
    open_eth_reg_write MAC[0x30] <- 0x00001001
        .
        .
        .
    open_eth_start_xmit TX: 0x3ffbb964, len: 42, tx_len: 42
    open_eth_desc_write DESC[0x0004] <- 0x3ffbb964
ethernet_connect: Got IPv6 event: Interface "example_netif_eth" address: fe80:0000:0000:0000:0000:00ff:fe12:3456, type: ESP_IP6_ADDR_IS_LINK_LOCAL
example_common: Connected to example_netif_eth
example_common: - IPv4 address: 10.0.2.15,
example_common: - IPv6 address: fe80:0000:0000:0000:0000:00ff:fe12:3456, type: ESP_IP6_ADDR_IS_LINK_LOCAL

example: Starting server on port: '80'
example: Registering URI handlers
    open_eth_reg_read MAC[0x30] -> 0x00001001
    open_eth_reg_write MAC[0x30] <- 0x00001001
        .
        .
        .
    open_eth_desc_read DESC[0x001c] -> 0x3ffbacdc
    open_eth_receive RX: len: 137
    open_eth_receive_desc RX: 0x3ffbacdc, len_flags: 0x008d4000
    open_eth_update_irq IRQ <- 0x4
    open_eth_reg_read MAC[0x04] -> 0x00000004
    open_eth_reg_write MAC[0x04] <- 0x00000004
    open_eth_update_irq IRQ <- 0x0
    open_eth_desc_read DESC[0x0024] -> 0x3ffbb320

example: Found header => Host: localhost:8000
    open_eth_desc_read DESC[0x0000] -> 0x003a2000
    open_eth_desc_write DESC[0x0000] <- 0x0074a000
    open_eth_start_xmit TX: 0x3ffbb964, len: 116, tx_len: 116
    open_eth_desc_write DESC[0x0004] <- 0x3ffbb964
    open_eth_receive RX: len: 60
    open_eth_receive_desc RX: 0x3ffbb320, len_flags: 0x00406000
    open_eth_update_irq IRQ <- 0x4
\end{lstlisting}

All lines with the ``open\_eth'' prefix represent functions calls to the OpenCores Ethernet MAC driver.  Lines 1-35 show the ESP32 load initializing the network stack through a sequence of calls to discover the underlying ethernet interface.  The OpenCores driver services these requests and passes through to the host, eventually responding back to the guest OS.  Reads and writes to the MAC register are handled via the driver, followed by data transmit (line 20), receive (line 21) and interrupt masking (line 23).

The next steps in lines 36-49 show the binding of the IP stack to the ethernet interface.  Again, the ethernet interface is serviced by the OpenCores driver utilizing the same open\_eth functions calls to read, write the MAC register and transmit data frames.

Lines 50-56 show the application HTTP server starting and binding port 80 with the IP stack.  As IP packet headers are stripped and forwarded to the MAC layer, the OpenCores driver services these requests and passes to the host.  The same API is used with function calls to read the MAC register (line 52) and write the MAC (line 53),

The final set is after we hit the host port 8000 with an HTTP request.  These are shown lines 57-73.  The host command

\begin{lstlisting}
    curl http://localhost:8000/hello
\end{lstlisting}

sends an HTTP (IP) request to the guest.  When this is broken down into ethernet frames, we see the OpenCores driver receive the request data (line 58) and set the interrupt mask (line 60).  The HTTP server then sends its response.  The server itself writes a message on line 66 that it has responded to the request from localhost:8000.  The data payload contains ``Hello World!'' and its transmit is handled by OpenCores on line 69.

\begin{figure}[h]
    \centerline{\includegraphics[width=1.0\linewidth, keepaspectratio]{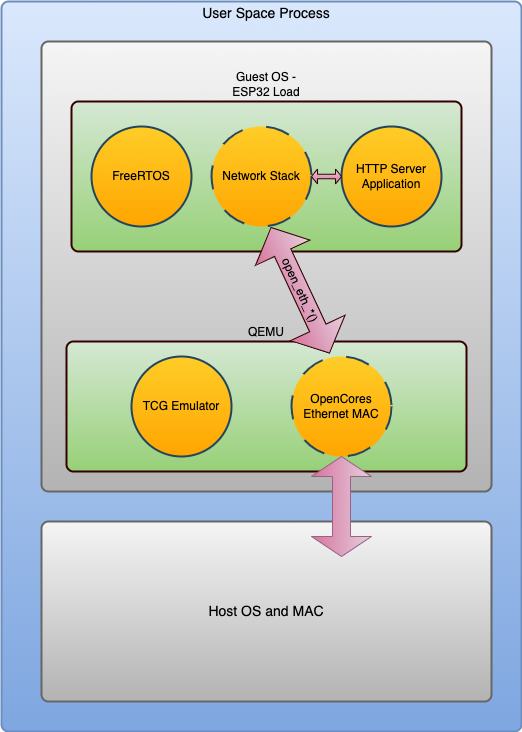}}
    \captionsetup{width=.8\linewidth}
    \caption{Components of the guest OS and the network data flow to the host. The Tiny Code Generator (TCG) emulates the Xtensa-native instructions while the OpenCores Ethernet MAC driver forwards ethernet traffic to the host MAC.}\label{fig:emulation_layers}   
\end{figure}

For a graphical representation of the above analysis, see Figure \ref{fig:emulation_layers}.

\newpage
\section{Future Work}\label{sec:future}
In this paper, we strive to emulate the ESP32 and understand how the medium access control (MAC) layer is handled within the emulation environment.  The evaluation is based on successfully passing network traffic between the host and emulated guest OSes in the form of HTTP request/response messages.  An important question to ask: how will this system behave at scale?

A logical next experiment is to evaluate the networking performance of the guest OS.  If we take the performance of the native ESP32 hardware as the gold standard, a comparison may be run to determine the potential latency of emulating the ESP32 instructions and the MAC layer.  A high rate of HTTP request messages would need to be sent to the target and the corresponding response delay measured.  In addition, with the containerizing of this emulation environment, an experiment may be performed to deploy numerous guest OS instances to evaluate how the response delay degrades as multiple emulated MAC instances communicate with the same host MAC.

\section{Conclusion}\label{sec:conclusion}
The goal of this project is to gain familiarity with the ESP32 microcontroller, build toolchain, SDK and explore possibilities to emulate and containerize.  Espressif maintains both the SDK (ESP-IDF) and a customized fork of the QEMU project on GitHub.  Additionally, Espressif publishes a VSCode extension which wraps the SDK and provides a convenient integrated development environment for coding.  An example HTTP server was easily built and linked with the SDK's suite of support libraries and FreeRTOS kernel.  It loads and runs without issue on the FreeNove test board purchased for these experiments.

The QEMU fork for ESP32 maintains pre-built binaries for Linux and can also be natively built for macOS.  Both options were explored.  The former was run through building a Docker container with the QEMU emulator and mounting the ESP32 load from the host file system at runtime.  The latter required building QEMU for macOS and launching with the ESP32 load.  Both options successfully emulated the Xtensa instruction set and allowed the ESP32 load to run without error.

While running on the FreeNove board, the ESP32 joined the available Wi-Fi for networking.  During emulation, it was configured to use the native ethernet interface.  This interface is provided by the OpenCores Ethernet MAC driver that is part of the QEMU project.  The driver provides an API to the guest OS network stack to discover the underlying MAC as well as transmit and receive ethernet frames.  These frames are forwarded between QEMU and the host OS.  QEMU has a trace framework built in with calls placed in all major API functions of every subsystem.  For these experiments, the OpenCores driver API activity was observed through launching QEMU with the option to trace all activity in the ``open\_eth'' subsystem.  Both the macOS and containerized Linux versions of QEMU ran the HTTP server networking application with no errors.

In conclusion, the ESP32 is a feature rich and cost-effective system-on-a-chip providing Wi-Fi, Bluetooth and a large selection of I/O devices built in to a tiny package.  The \$10 investment provides a powerful microcontroller that is ideal for car chargers, stand mixers, robotic joints and many other embedded applications.  It is possible to natively build the ESP32 and emulate this load through QEMU.  Containerizing this combination adds a very convenient way to improve the rapid build and test cycle while scaling to many device deployments on a single host. 

\newpage
\section*{Acknowledgements}
The author would like to thank Dr Bruce Irvin at Portland State University for the fantastic course he taught in the concepts of operating systems as well as his valuable advice, direction and mentorship.

\bibliography{IEEEabrv,\docroot/References/References}

\end{document}